\documentclass[12pt]{iopart}

\usepackage{iopams}  
\usepackage[pdftex]{graphicx}
\usepackage{setspace,bm,color}
\usepackage[colorlinks=true, pdfstartview=FitV, linkcolor=red, citecolor=blue, urlcolor=blue]{hyperref}
\usepackage[caption=false]{subfig}
\usepackage{slashed}
\usepackage[numbers,sort&compress]{natbib}
\graphicspath{{./Figures/}}

\newcommand{\diff}{\mathrm{d}}
\newcommand{\p}{\partial}
\newcommand{\ve}{\varepsilon}

\newcommand{\up}{\uparrow}
\newcommand{\down}{\downarrow}
\newcommand{\be}{\begin{equation}}      
\newcommand{\ee}{\end{equation}}      
\newcommand{\bea}{\begin{eqnarray}}      
\newcommand{\eea}{\end{eqnarray}}

\newcommand{\im}{\mathrm{i}}

\newcommand{\eqref}[1]{(\ref{#1})}

\begin{document}

\title{Lefschetz-thimble analysis of the sign problem in one-site fermion model}

\author{Yuya Tanizaki$^{1,2}$, Yoshimasa Hidaka$^2$ and Tomoya Hayata$^3$}

\address{$^1$ Department of Physics, The University of Tokyo, Tokyo 113-0033, Japan}
\address{$^2$ Theoretical Research Division, Nishina Center, RIKEN, Wako, Saitama 351-0198, Japan}
\address{$^3$ RIKEN Center for Emergent Matter Science (CEMS), Wako, Saitama 351-0198, Japan}
\eads{\mailto{yuya.tanizaki@riken.jp}, \mailto{hidaka@riken.jp}, \mailto{hayata@riken.jp}}

\vspace{10pt}
\begin{indented}
\item[]\today
\end{indented}

\begin{abstract}
The Lefschetz-thimble approach to path integrals is applied to a one-site model of electrons, i.e., the one-site Hubbard model. 
Since the one-site Hubbard model shows a non-analytic behavior at the zero temperature 
and its path integral expression has the sign problem, this toy model is a good testing ground for an idea or a technique to attack the sign problem.  
Semiclassical analysis using complex saddle points unveils the significance of interference among multiple Lefschetz thimbles to reproduce the non-analytic behavior by using the path integral. 
If the number of Lefschetz thimbles is insufficient, we found not only large discrepancies from the exact result, but also thermodynamic instabilities. 
Analyzing such singular behaviors semiclassically, we propose a criterion to identify the necessary number of Lefschetz thimbles. 
We argue that this interference of multiple saddle points is a key issue to understand the sign problem of the finite-density quantum chromodynamics. 
\end{abstract}

% Uncomment for PACS numbers
%
% Uncomment for keywords
%\vspace{2pc}
%\noindent{\it Keywords}: XXXXXX, YYYYYYYY, ZZZZZZZZZ
%
% Uncomment for Submitted to journal title message
%\submitto{\JPA}
%
% Uncomment if a separate title page is required
%\maketitle
% 
% For two-column output uncomment the next line and choose [10pt] rather than [12pt] in the \documentclass declaration
%\ioptwocol
%
\section{Introduction}\label{sec:intro}

%High-temperature superconductivity \cite{Anderson1987Science}. 

Understanding strongly-correlated quantum many-body systems has been an ultimate goal in contemporary physics.
Numerical simulation formulated on the discretized spacetime, especially lattice quantum Monte Carlo method, is a powerful \textit{ab initio} tool for that purpose. 
Exact diagonalization of a Hamiltonian provides us complete information on the system; 
however, it usually requires the huge computational cost and is limited to small systems. 
Monte Carlo method, on the other hand, is based on the importance sampling in the phase space of the system, 
whose computational cost scales algebraically with the system size. 
Many thermodynamic quantities can be computed for various systems using this method, such as finite-temperature quantum chromodynamics (QCD) in hadron physics~\cite{Karsch:2001cy}, and liquid helium~\cite{RevModPhys.67.279}, ultracold atomic gases~\cite{Pollet2012}, Bose--Fermi mixtures~\cite{Yamamoto:2012wy} in condensed matter physics.    

In many quantum systems of great interest, however, Monte Carlo simulation suffers from the so-called sign problem~\cite{PhysRevB.41.9301, Batrouni:1992fj}. 
The path-integral weight $\mathrm{e}^{-S}$ becomes negative or even complex, and thus the importance sampling cannot be applied. 
Since the cancellation between large positive and negative noises gives a physically meaningful signal, the computational time to reduce statistical errors grows exponentially with the system size~\cite{Troyer:2004ge}. 
In condensed matter physics, famous examples of the sign problem are strongly-correlated electrons, such as the Hubbard model away from half-filling~\cite{PhysRevB.41.9301}, frustrated spin systems, such as the XY model on the Kagom\'{e} lattice, and so on.
In hadron physics, QCD at finite quark densities attracts much attention for understanding the interior structure of neutron stars~\cite{RevModPhys.80.1455, Fukushima:2010bq, Masuda:2012kf}, but we have no \textit{ab initio} simulation due to the sign problem~\cite{Muroya:2003qs}. 

The sign problem of the finite-density QCD is known to become too severe when the quark chemical potential exceeds half of the pion mass \cite{Barbour:1997bh,Barbour:1997ej, Cohen:2003kd, Adams:2004yy, Hidaka:2011jj, Hanada:2012es}. 
At low temperatures, there is no phase transition until the quark chemical potential reaches about one third of the nucleon mass; however, the Monte Carlo simulation of two flavor QCD shows a phase transition at half of the pion mass~\cite{Barbour:1997bh,Barbour:1997ej}. No one can describe the true onset at one third of the nucleon mass by using the path-integral formulation of lattice QCD \cite{Cohen:2003kd, Adams:2004yy}. 
This is one of the biggest issue in finite-density QCD.
%hadron physics when we try to understand cold and dense nuclear matters realized inside neutron stars. 
In this paper, we address this problem, which is called the ``Silver Blaze problem"~\cite{Cohen:2003kd}, by studying a toy model. 

There have been many attempts to attack the sign problem. %in order to circumvent this big obstacle. 
The idea of complexification of the integration variables is recently developing, and the complex Langevin method~\cite{Klauder:1983zm, PhysRevA.29.2036, Parisi:1984cs, Damgaard:1987rr} and the Lefschetz-thimble approach~\cite{pham1983vanishing,Witten:2010cx, Witten:2010zr, Harlow:2011ny} to the path integral rely on this idea. 
The Picard--Lefschetz theory gives an extension of the steepest descent method to multiple oscillatory integrals, and Lefschetz thimbles correspond to steepest descent paths in the complex contour integration~\cite{pham1983vanishing}. 
After finding its usefulness in studying analytic properties of the Chern--Simons theory with the semiclassical analysis~\cite{Witten:2010cx, Witten:2010zr, Harlow:2011ny}, 
it leads to diverse applications to real-time quantum phenomena~\cite{Tanizaki:2014xba, Cherman:2014sba},  resurgence trans-series methods of perturbation theories~\cite{Unsal:2012zj, Basar:2013eka, Cherman:2014ofa, Dorigoni:2014hea, Behtash:2015kna, Behtash:2015kva}, and the sign problem in lattice simulations~\cite{Mukherjee:2014hsa, Cristoforetti:2012su, Cristoforetti:2013wha, Cristoforetti:2014gsa, Aarts:2013fpa, Fujii:2013sra, Aarts:2014nxa, DiRenzo:2015foa, Fukushima:2015qza}. 
The Lefschetz-thimble approach is based on rigorous mathematics; however, its practical properties have not been fully understood yet. 
For recent developments, see Refs.~\cite{Tanizaki:2014tua, Kanazawa:2014qma, Tanizaki:2015pua}.

In this paper, we apply the Lefschetz-thimble approach to the one-site model of electrons. 
This toy model can be regarded as an extreme limit of strong couplings because it can be obtained by neglecting hopping terms in the Hubbard model. The Hamiltonian of the one-site model can be easily diagonalized and thus we can calculate any expectation value exactly. 
However, since this model has the severe sign problem in its path-integral expression, it is hard to calculate expectation values by the conventional Monte Carlo method. 
This toy model provides us a good playground to study theoretical structures of the Lefschetz-thimble approach. % for an application to the sign problem. 
In a previous study~\cite{Mukherjee:2014hsa}, two-dimensional Hubbard model is studied using the Lefschetz-thimble Monte Carlo method, but one must use the so-called ``one-thimble approximation" due to the current limitation of the numerical algorithm. 
Since our system is exactly solvable, we can understand the complete structure of Lefschetz thimbles, and suggest appropriate approximation schemes. 

Based on the semiclassical analysis, we find that interference of complex phases among multiple saddle points is important to reproduce the 
step-function behavior of the density. 
The fermion spectrum at a complex saddle point resembles the quark spectrum of phase quenched finite-density QCD, which would suggest the interference of multiple saddle points might also occur at finite-density QCD. We will discuss this point in detail later.
This study will help us to understand the sign problem in QCD. 

The outline of this paper is as follows. 
In Sec.~\ref{sec:HubbardModel}, the one-site Hubbard model is explained, and its physical properties are calculated by exactly diagonalizing the Hamiltonian. 
The path integral expression of the toy model is introduced in detail, and the sign problem appearing in its path integral formulation is briefly reviewed. 
In Sec.~\ref{sec:LefschetzThimble_Lefschetz-thimble}, the Lefschetz-thimble method is introduced, and is applied to the one-site Hubbard model. 
Complex saddle points play a pivotal role in this approach, and thus we study their property in Sec.~\ref{sec:SemiClassical_OneSiteHubbard}. 
We discuss structures of Lee--Yang zeros and the fermion spectrum at complex saddle points in Sec.~\ref{sec:LeeYangHubbard}. 
In Sec.~\ref{sec:NumericalResults_OneSiteHubbard}, numerical results on the fermion number density are compared with the exact result using several approximations based on the Lefschetz-thimble approach. %, and they are analyzed based on complex saddle-point analysis. 
%We observed that interference among multiple Lefschetz thimbles has a significant effect to explain the quantum phase transition of this model. 
In Sec.~\ref{sec:speculation_silverblaze_qcd}, we discuss the resemblance of this model to the finite-density QCD and address the Silver Blaze problem in QCD. 
We summarize our result in Sec.~\ref{sec:summary}. 

%----------------------------------------------------------------
%Hubbard Model
%----------------------------------------------------------------
\section{One-site Hubbard model}\label{sec:HubbardModel}
\subsection{Physical properties of the model}\label{sec:HubbardModel_properties}
The Hubbard model~\cite{Hubbard238, PhysRevLett.10.159, Kanamori1963} describes the lattice fermion system, whose dynamics is governed by the (second quantized) Hamiltonian,
\bea
\hat{H}_{\rm F}&=&-t\sum_{\langle i,j \rangle,\sigma}\hat{c}^{\dagger}_{\sigma,i} \hat{c}_{\sigma,j}+U \sum_{i}\hat{n}_{\up,i}\hat{n}_{\down,i}-\mu\sum_{i}(\hat{n}_{\up,i}+\hat{n}_{\down,i}). 
\label{Eq:Hamiltonian_Hubbard}
\eea
The summation is taken over all the lattice sites $i$, and the notation $\langle i,j\rangle$ denotes that only the nearest neighbor hopping is considered. 
$\hat{c}^{\dagger}_{\sigma,i}$and $\hat{c}_{\sigma,i}$, obeying $\left\{\hat{c}_{\sigma,i}, \hat{c}_{\tau,j}^{\dagger}\right\}=\delta_{\sigma\tau}\delta_{ij}$, are creation and annihilation operators of fermions of the spin $\sigma(=\up,\down)$ at the site $i$. $\hat{n}_{\sigma,i}=\hat{c}^{\dagger}_{\sigma,i}\hat{c}_{\sigma,i}$ is the number operator. 
The parameter $t$ is called the hopping parameter, and we will consider only the case $t=0$ in this section. 
The parameter $U(>0)$ describes the on-site repulsive interaction: The energy increases if two particles get together on the same site. 
The Hamiltonian is invariant under the global $U(1)$ rotation $\hat{c}_{\sigma,i}^{(\dagger)}\mapsto \exp(\pm\im \alpha)\hat{c}_{\sigma,i}^{(\dagger)}$, and thus the total number of fermions is conserved. 
In the Hamiltonian \eqref{Eq:Hamiltonian_Hubbard}, the chemical potential $\mu$ is introduced for this conserved quantity. 

In the strong coupling limit $t=0$, one can solve the Hamiltonian \eqref{Eq:Hamiltonian_Hubbard} analytically. Let us see the result on the total number density $\langle \hat{n}\rangle:=\langle (\hat{n}_{\up}+\hat{n}_{\down})\rangle$ as a function of the chemical potential $\mu$. 
Since each site is totally independent from others, one can study the one-site Hamiltonian,
\be
\hat{H}={U}\hat{n}_{\up}\hat{n}_{\down}-\mu\hat{n}. 
\label{Eq:Hamiltonian_SS_Hubbard}
\ee
In the following, we only consider the one-site case. 
Since the Hamiltonian \eqref{Eq:Hamiltonian_SS_Hubbard} commutes with the number density operator  $\hat{n}$, we can take the number basis to find the ground state. 
Let us define the grand partition function by 
\be
Z=\mathrm{tr}\left[\exp-\beta \hat{H}\right], 
\label{Eq:PartitionFunction_Hubbard}
\ee
with the inverse temperature $\beta=1/T$.
The number basis gives an explicit result: 
\be
Z=1+2\mathrm{e}^{\beta\mu}+\mathrm{e}^{\beta(2\mu-U)} .
\label{Eq:Result_PartitionFunction_Hubbard}
\ee
The number density is given as 
\be
\langle \hat{n}\rangle={1\over \beta}{\p\over \p \mu}\ln Z ={2(\mathrm{e}^{\beta\mu}+\mathrm{e}^{\beta(2\mu-U)})\over 1+2\mathrm{e}^{\beta\mu}+\mathrm{e}^{\beta(2\mu-U)}}.
\label{Eq:NumberDensity_Hubbard}
\ee
In the zero-temperature limit $\beta\to\infty$, the number density takes $0$, $1$, and $2$ for $\mu/U<0$, $0<\mu/U<1$, and $\mu/U>1$, respectively, and shows the step-function like behavior. 
In particular in the case $\mu/U=1/2$, we have $\langle \hat{n}\rangle=1$ for any $\beta$, which comes from the symmetry $\hat{n}\mapsto 2-\hat{n}$ (or, $\hat{n}_{\sigma}\mapsto 1-\hat{n}_{\sigma}$) in the Hamiltonian~\eqref{Eq:Hamiltonian_SS_Hubbard}. 
This point is called half-filling, at which the particle-hole symmetry exists.

%--------------------------------------------------------------
\subsection{Path integral formulation and the sign problem}\label{sec:HubbardModel_pathintegral}
We have explicitly seen that the one-site Hubbard model can be analytically solved  by using the number eigenstates. 
However, if one changes the basis of the Hilbert space for taking trace, the sign problem emerges and thus the Hubbard model in the strong coupling provides us a good lesson. 

Let us first derive the path integral expression of the partition function~\eqref{Eq:PartitionFunction_Hubbard}. 
Using two-component complex Grassmannian variables $\psi=(\psi_{\up},\psi_{\down})$, the fermion coherent state is defined by 
\be
|\psi\rangle=\exp(-\psi \hat{c}^{\dagger})|0\rangle, 
\ee
which satisfies $\hat{c}_{\sigma}|\psi\rangle=\psi_{\sigma}|\psi\rangle$. 
We take the convention in which all the Grassmannian variables anticommute with each other and with fermionic creation/annihilation operators. 
It is easy to check that 
\be
\langle \psi_1|\psi_2\rangle=\mathrm{e}^{\psi_1^*\psi_2},\;\quad 
\int\diff \psi^*\diff \psi |\psi\rangle \mathrm{e}^{-\psi^*\psi}\langle \psi|=1, 
\ee
and 
\be
\mathrm{tr}(O)=\int\diff \psi^*\diff \psi \,\mathrm{e}^{-\psi^*\psi}\langle -\psi|O|\psi\rangle. 
\ee
Let $|n_{\up},n_{\down}\rangle$ be a fermionic Fock state, and then the inner products with coherent states are 
\bea
\langle 0,0|\psi\rangle=1,\; \langle 1,0|\psi\rangle=\psi_{\up},\; \langle 0,1|\psi\rangle=\psi_{\down},\; 
\langle 1,1|\psi\rangle=-\psi_{\up} \psi_{\down}. 
\eea
Using these relations, one can compute the matrix element of the imaginary-time evolution operator as 
\begin{eqnarray}
\mathrm{e}^{-\psi^*\psi}\langle \psi | \mathrm{e}^{-\Delta \tau \hat{H}}|\psi'\rangle 
\simeq \exp-\Bigl[\psi^*\left(\psi-\mathrm{e}^{\Delta \tau \mu}\psi'\right)%\nonumber\\&\quad
+{\Delta \tau\over 2} U(\psi^*\mathrm{e}^{\Delta \tau \mu}\psi')^2\Biggr]. 
\label{Eq:MatrixElement_Hubbard_01}
\end{eqnarray}
Here we take an approximation for small $\Delta \tau$. 
In order to circumvent the quartic interaction, let us introduce an auxiliary real field $\varphi$ via the Hubbard--Stratonovich transformation, that is, 
\be
1=\sqrt{\Delta \tau\over 2\pi U}\int\diff \varphi\exp-{\Delta \tau\over 2U}
 \Bigl(\varphi-\im  U(\psi^*\mathrm{e}^{\Delta\tau \mu}\psi{'})\Bigr)^2. 
\label{Eq:AuxiliaryField_Hubbard}
\ee
Inserting Eq.~\eqref{Eq:AuxiliaryField_Hubbard} into Eq.~\eqref{Eq:MatrixElement_Hubbard_01}, we obtain 
\begin{eqnarray}
&\mathrm{e}^{-\psi^*\psi}\langle \psi | \mathrm{e}^{-\Delta \tau \hat{H}}|\psi'\rangle \nonumber\\
&\simeq\sqrt{\Delta \tau\over 2\pi U}\int{\diff \varphi}\exp-\Biggl({\Delta \tau\over 2U}\varphi^2%\nonumber\\
+\psi^*\left(\psi-\mathrm{e}^{\Delta \tau \mu}(1+\im \Delta \tau \varphi)\psi'\right)\Biggr). 
\label{Eq:MatrixElement_Hubbard_02}
\end{eqnarray}
In order to take the naive continuum limit of the path integral expression, we need to exponentiate the auxiliary field in the fermionic bilinear term as 
\begin{eqnarray}
\mathrm{e}^{\Delta \tau \mu}(1+\im \Delta \tau \varphi)
&=\exp\left(\Delta\tau (\mu+\im \varphi)+{\Delta \tau^2\over 2}\langle\varphi^2\rangle_0+O(\Delta \tau^{3/2}) \right)
\nonumber\\
&=\exp\left(\Delta\tau \left(\mu+{U\over 2}+\im \varphi\right)+O(\Delta \tau^{3/2}) \right). 
\end{eqnarray}
Here, $\langle\varphi^2\rangle_0$ means the expectation value of $\varphi^2$ with the Gaussian weight in Eq.~\eqref{Eq:MatrixElement_Hubbard_02}, which gives $U/\Delta\tau$. The path integral expression of the partition function reads 
\begin{eqnarray}
Z&=\lim_{N\to \infty}\sqrt{\beta/N\over 2\pi U}^N\int \prod_{k=1}^N\diff \varphi_k \exp\left(-{\beta\over N}\sum_{k=1}^{N}{\varphi_k^2\over 2U}\right)  \nonumber\\ 
&\quad\times \int\prod_{k=1}^{N}\diff \psi_k^*\diff \psi_k\exp-\left(\sum_{k=1}^{N} \psi_{k+1}^*\left(\psi_{k+1}-\mathrm{e}^{{\beta\over N}(\im\varphi_k +\mu+U/2)}\psi_k\right)\right), 
%Z&=\int \Diff \varphi \exp\left(-\int_0^{\beta}\diff \tau {\varphi^2\over 2U}\right)  \nonumber\\ 
%&\times \int\Diff \psi^*\Diff \psi\exp\left(-\int_0^{\beta}\diff \tau\psi^*\left(\p_{\tau}-\im\varphi -\mu-U/2\right)\psi\right).
\label{Eq:PathIntegral_Hubbard_01}
\end{eqnarray}
with the antiperiodic boundary condition $\psi_{N+1}=-\psi_{1}$. 
%with $\mu'=\mu+U/2$. 
Equation of motion in terms of $\varphi$ shows 
\be
\langle \hat{n}\rangle=\langle \psi^*_{k+1}\mathrm{e}^{{\beta\over N}(\im\varphi_k+\mu+U/2)}\psi_k\rangle = -{\im\over U}\langle \varphi_k\rangle . 
\ee
The expectation value of $\varphi$ is nothing but that of the total number density. 
This formula will be used later in order to compute the number density. 
Since Eq.~\eqref{Eq:PathIntegral_Hubbard_01} is quadratic in the fermionic field $\psi$ and $\psi^*$, we can perform the Grassmannian integration explicitly. 
The spectrum of the fermion bilinear operator in the exponential in Eq.~\eqref{Eq:PathIntegral_Hubbard_01} is given by 
\be
\lambda_\ell(\varphi,\mu)=1-\mathrm{e}^{(2\ell-1)\pi \im/N}\exp {\beta\over N^2}\sum_{k=1}^{N}(\im \varphi_k+\mu+U/2),
\label{eq:FermionSpectrumHubbard}
\ee
and then, in the continuum limit, the fermion path integral (determinant) becomes
\be 
\left(\prod_{\ell=1}^{N}\lambda_\ell\right)^2=
\left(1+\exp\int_0^\beta\diff\tau\left(\im \varphi(\tau)+\mu+U/2\right)\right)^2. 
\label{Eq:FermionicDeterminantHubbard}
\ee
Since fermions have two spin degrees of freedom, the overall square is taken. 
The fermion determinant \eqref{Eq:FermionicDeterminantHubbard} contains only the Matsubara zero mode of $\varphi$. 
Thus the path integral of non-zero Matsubara modes of $\varphi$  gives a trivial Gaussian integration and does not depend on $\mu$. 
The path integral of our interest is now reduced to an integral of zero Matsubara mode $\varphi_{\mathrm{bg}}=\int_0^\beta\diff\tau\varphi(\tau)/\beta$, and we have 
\be
Z=\sqrt{\beta\over 2\pi U}\int {\diff \varphi_{\mathrm{bg}}}  \left(1+\mathrm{e}^{\beta\left(\im \varphi_{\mathrm{bg}}+\mu+U/2\right)}\right)^2\mathrm{e}^{-{\beta\varphi_{\mathrm{bg}}^2/ 2U}}. 
\label{Eq:ZeroModeIntegral_Hubbard}
\ee
This integral can be performed analytically to find Eq.~\eqref{Eq:Result_PartitionFunction_Hubbard}. 
Instead, we shall apply the Lefschetz-thimble method because Eq.~\eqref{Eq:ZeroModeIntegral_Hubbard} has the sign problem coming from the complex weight $\mathrm{e}^{\im\beta\varphi_{\mathrm{bg}}}$. 
This simple model enables us to study the structure of the sign problem in  terms of the Lefschetz-thimble approach. 

In order to explicitly show that the path integral \eqref{Eq:ZeroModeIntegral_Hubbard} contains the sign problem, let us compare it with the phase quenched partition function 
\be
Z'=\sqrt{\beta\over 2\pi U}\int {\diff \varphi_{\mathrm{bg}} }  \left|1+\mathrm{e}^{\beta\left(\im \varphi_{\mathrm{bg}}+\mu+U/2\right)}\right|^2\mathrm{e}^{-{\beta}\varphi_{\mathrm{bg}}^2/2U}.
\label{Eq:QuenchedZ_woShift}
\ee
This integration can be performed analytically as
\begin{equation}
Z'=1+2\mathrm{e}^{\beta\mu}+\mathrm{e}^{\beta(2\mu+U)}.
\label{eq:Z'}
\end{equation}
The positive sign in front of $U$ is different from that in Eq.~\eqref{Eq:Result_PartitionFunction_Hubbard}.
If the last term is negligible, the sign problem is mild. This is the case for small chemical potentials $\mu/U \lesssim -1/2$ and low temperatures $\beta U\gg 1$; the ratio $Z/Z'$ is almost one because the dominant contribution comes from the unoccupied state. Therefore the sign problem is sufficiently weak. 
In the later section, we can interpret this semiclassically because the important configurations accumulate around the origin $\varphi_{\mathrm{bg}}\simeq 0$. 
The sign problem becomes quite severe for $\mu/U \gtrsim -1/2$ and low temperatures $\beta U\gg 1$. For example, at $\mu/U =0$,  the ratio becomes 
\be
Z/Z'\simeq 3\,\mathrm{e}^{-\beta U}\ll 1. 
\ee
This severe sign problem comes from the last term in Eq.~\eqref{eq:Z'}, which causes the wrong onset of the density at  $\mu/U=-1/2$. A similar problem happens in QCD at finite density when the quark chemical potential exceeds the half of the pion mass~\cite{Barbour:1991vs,Barbour:1997bh,Barbour:1997ej}.
For large chemical potentials, $\mu/U\gtrsim 3/2$, we can switch the occupied and unoccupied fermions by changing the integration variable from $\varphi_{\mathrm{bg}}$ to $-\varphi_{\mathrm{bg}}+2\im U$, so that the sign problem is weakened. 
There is another remarkable condition in which the sign problem disappears again by a simple shift of the integration variable: 
At the half-filling $\mu/U=1/2$, a new integration variable $\varphi_{\mathrm{bg}}=\varphi'_{\mathrm{bg}}+\im U$ rewrites the original integral \eqref{Eq:ZeroModeIntegral_Hubbard} into 
\be
Z=\sqrt{8\beta\over \pi U}\int {\diff \varphi'_{\mathrm{bg}} }\left(\cos{\beta \varphi'_{\mathrm{bg}}\over 2}\right)^2 \exp-{\beta\over 2U}({\varphi'_{\mathrm{bg}}}^2-U^2). 
\label{Eq:HalffillingIntegral_Hubbard}
\ee
The integrand of Eq.~\eqref{Eq:HalffillingIntegral_Hubbard} is nonnegative. 
This is because the particle-hole symmetry at the half-filling becomes manifest in this new integration variable. 
As a conclusion of this section, the sign problem can be circumvented in these three special cases, $\mu/U\lesssim-1/2$, $\mu/U\gtrsim3/2$ and $\mu=1/2$, even using the path integral expression; however, this does not happen in general cases. 

\section{Lefschetz-thimble approach} %to the repulsive Hubbard model}
\label{sec:LefschetzThimble_OneSiteHubbard}

\subsection{Lefschetz-thimble method}\label{sec:LefschetzThimble_Lefschetz-thimble}
Let us start with a multiple integration that gives the partition function
\be
Z=\int_{\mathbb{R}^n}\diff^n x~\mathrm{e}^{-S(x)},
\label{Eq:General_Expression_Partition_Function}
\ee
where $S(x)$ is a complex action functional of the real field $x=(x^1,\ldots,x^n)\in\mathbb{R}^n$. 
In order to circumvent the oscillatory integral~\eqref{Eq:General_Expression_Partition_Function}, we complexify the integration variables $x^j\mapsto z^j=x^j+\im y^j$\footnote{There is another approach to the sign problem also based on the idea of complexification, 
the complex Langevin method~\cite{Parisi:1984cs, Damgaard:1987rr}. See Refs.~\cite{Aarts:2010aq, Aarts:2011ax, Anzaki:2014hba, Makino:2015ooa,Nishimura:2015pba} for recent developments. 
}. 
Integration is performed on steepest descent paths, called Lefschetz thimbles, instead of directly evaluating Eq.~\eqref{Eq:General_Expression_Partition_Function}. 
Each Lefschetz thimble is an $n$-dimensional space spanned around a saddle point $z_{\sigma}$ in $\mathbb{C}^n$ ($\sigma\in\Sigma$). 
Let us consider Morse's flow equation for complexified variables $z$ by introducing a fictitious time $t$~\cite{pham1983vanishing, Witten:2010cx, Witten:2010zr}:
\begin{equation}
\frac{\diff {z^i}(t)}{\diff t} =   \overline{\left(\frac{\partial S(z(t))}{\partial z^i}\right)}.
\label{Eq:Downward_Flow}
\end{equation}
Along this flow, the integrand becomes smaller since 
\be
{\diff\over \diff t} \mathrm{Re}\; S(z(t))=\left|{\p S(z(t))\over \p z^i}\right|^2\ge 0, 
\ee
and thus $\exp-S(z(t))\to 0$ as $t\to\infty$. 
This flow equation has an important conserved quantity: The imaginary part of $S$ obeys 
\be
{\diff \over \diff t}\mathrm{Im}\, S(z(t))=0. 
\ee
Therefore, along each flow line, the integrand has no oscillatory phase, which is a remarkable property to circumvent the oscillatory integral. 
Now, let us define a new integration cycle called Lefschetz thimble as follows: 
The Lefschetz thimble $\mathcal{J}_{\sigma}$ is identified as the set
of points reached by some flows emanating from $z_{\sigma}$:
\be
\mathcal{J}_{\sigma}:=\{z(0)\,|\, z(-\infty)=z_{\sigma}\}. 
\ee
Complex contour integral over $\mathcal{J}_{\sigma}$ definitely converges since $\mathrm{Re}\, S$ diverges at boundaries of $\mathcal{J}_{\sigma}$. 
On each integral, we have no sign problem because $\mathrm{Im}\, S$ is constant. 
The partition function can now be computed by the sum of the nicely converging integrals as
\be
Z=\sum_{\sigma\in\Sigma}n_{\sigma}\int_{\mathcal{J}_{\sigma}}\diff^n z~\mathrm{e}^{-S(z)}.
\label{Eq:Lefschetz_Thimble_Decomposition}
\ee
The coefficient $n_{\sigma}$ is  given by the intersection number between $\mathbb{R}^n$ and $\mathcal{K}_{\sigma}$.
The dual thimble $\mathcal{K}_{\sigma}$ is defined as the set of the points reached by
flows getting sucked into $z_\sigma$:
\be
\mathcal{K}_{\sigma}=\{z(0)\, |\, z(+\infty)=z_{\sigma}\}. 
\ee 

There are two possible origins where the sign problem reappears in the Lefschetz-thimble method.
One is a complex phase coming from the Jacobian of the integration measure $\diff^n z$ on Lefschetz thimbles, which is often called the residual sign problem \cite{Aarts:2013fpa,
Aarts:2014nxa, Cristoforetti:2014gsa,  Fujii:2013sra}. 
Geometrically, the residual sign problem is mild if the Lefschetz thimble is almost straight. 
Another one comes from the possible cancellation in summing up Lefschetz thimbles in Eq.~\eqref{Eq:Lefschetz_Thimble_Decomposition}, because the complex phase $\mathrm{Im}\, S$ can be different among several Lefschetz thimbles. 

\subsection{Semiclassical analysis}\label{sec:SemiClassical_OneSiteHubbard}
Let us apply the Picard--Lefschetz theory to the path integral~\eqref{Eq:ZeroModeIntegral_Hubbard}. The effective action of this system is  
\be
S(z)={\beta\over 2U}z^2-2\ln \left(1+\exp\beta\left(\im z  +\mu+{U\over 2}\right)\right). 
\label{Eq:EffectiveAction_Hubbard}
\ee
The effective action satisfies the reality condition
\be
\overline{S(z)}=S(-\overline{z}), 
\label{Eq:RealityCondition}
\ee
and then the Lefschetz-thimble decomposition manifestly respects this symmetry so as to ensure the reality of physical observables \cite{Tanizaki:2015pua, Fukushima:2006uv, Dumitru:2005ng, Nishimura:2014rxa}. 

The logarithmic function has branch singularities at fermionic Matsubara modes $z=\im (\mu+U/2)+(2\ell+1)\pi /\beta$ for $\ell \in \mathbb{Z}$. 
The integrand converges to $0$, i.e., $S\to +\infty$, at these points. 
The flow equation of Eq.~\eqref{Eq:EffectiveAction_Hubbard} reads 
\be
{\diff z \over \diff t}={\beta\over U}\overline{z}+{2\im \beta\exp\beta\left(-\im\overline{z}+\mu+{U\over2}\right)\over 1+\exp\beta\left(-\im\overline{z}+\mu+{U\over2}\right)} . 
\label{Eq:Hubbard_DownwardFlow}
\ee
Let us find the set of saddle points, which is an important step not only for the saddle-point approximation but also for the Lefschetz-thimble method.  
The saddle-point condition of the effective action~\eqref{Eq:EffectiveAction_Hubbard} is 
\be
\im z_{\sigma}=-{2U\over 1+\exp-\beta\left(\im z_{\sigma}+\mu+{U\over2}\right)}. 
\ee
In order to simplify our analysis, we consider a limiting case where $T\ll U,|\mu|$. 
We can approximately obtain the saddle points as 
\be
z_{m}=\im\left(\mu+{U\over 2}\right)+T\left(2\pi m +\im\ln {{3\over2}U-\mu\over {1\over 2}U+\mu}\right)+O(T^2)
\label{Eq:ApproximateSaddlePoints}
\ee 
for $m\in\mathbb{Z}$ by assuming that the second term is much smaller than the first one. 
The reality condition \eqref{Eq:RealityCondition} says that $z_{-m}$ and $z_{m}$ form a pair. If $\mu>3U/2$ or $\mu<-U/2$, there is another solution 
\be
z_{*}=
\left\{\begin{array}{ccc}
2\im U+o(T) & {\rm for} & \mu/U>{3\over 2},\\ 
0+o(T) & {\rm for} & \mu/U<-{1\over 2}.
\end{array}\right. 
\ee

\begin{figure}[t]
\centering
\begin{minipage}{.43\textwidth}
\subfloat[$\mu/U=2$]{
\includegraphics[scale=0.42]{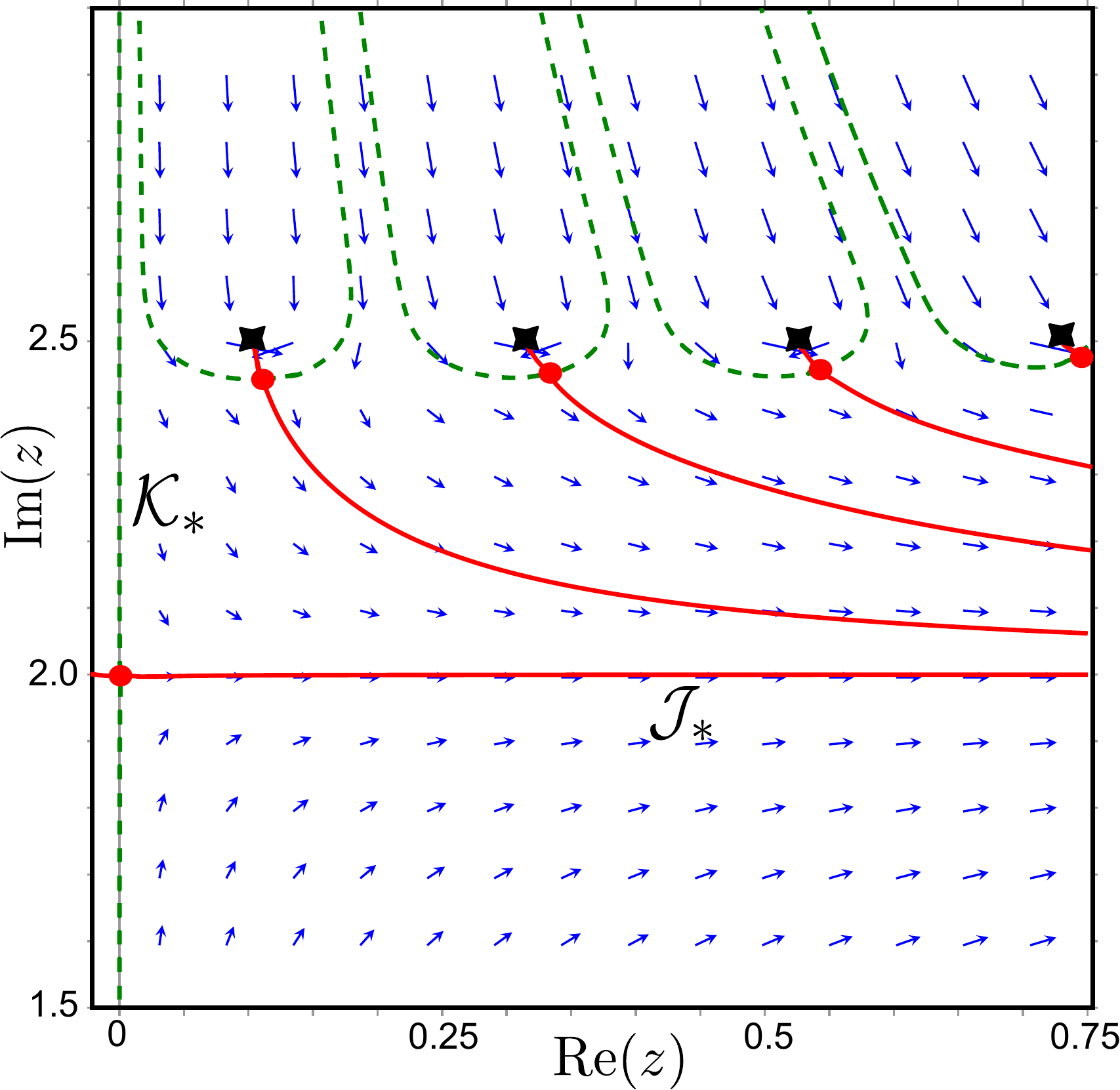}
}
\end{minipage}\qquad%\vspace{0.5em}
\begin{minipage}{.43\textwidth}
\subfloat[$\mu/U=0$]{
\includegraphics[scale=0.42]{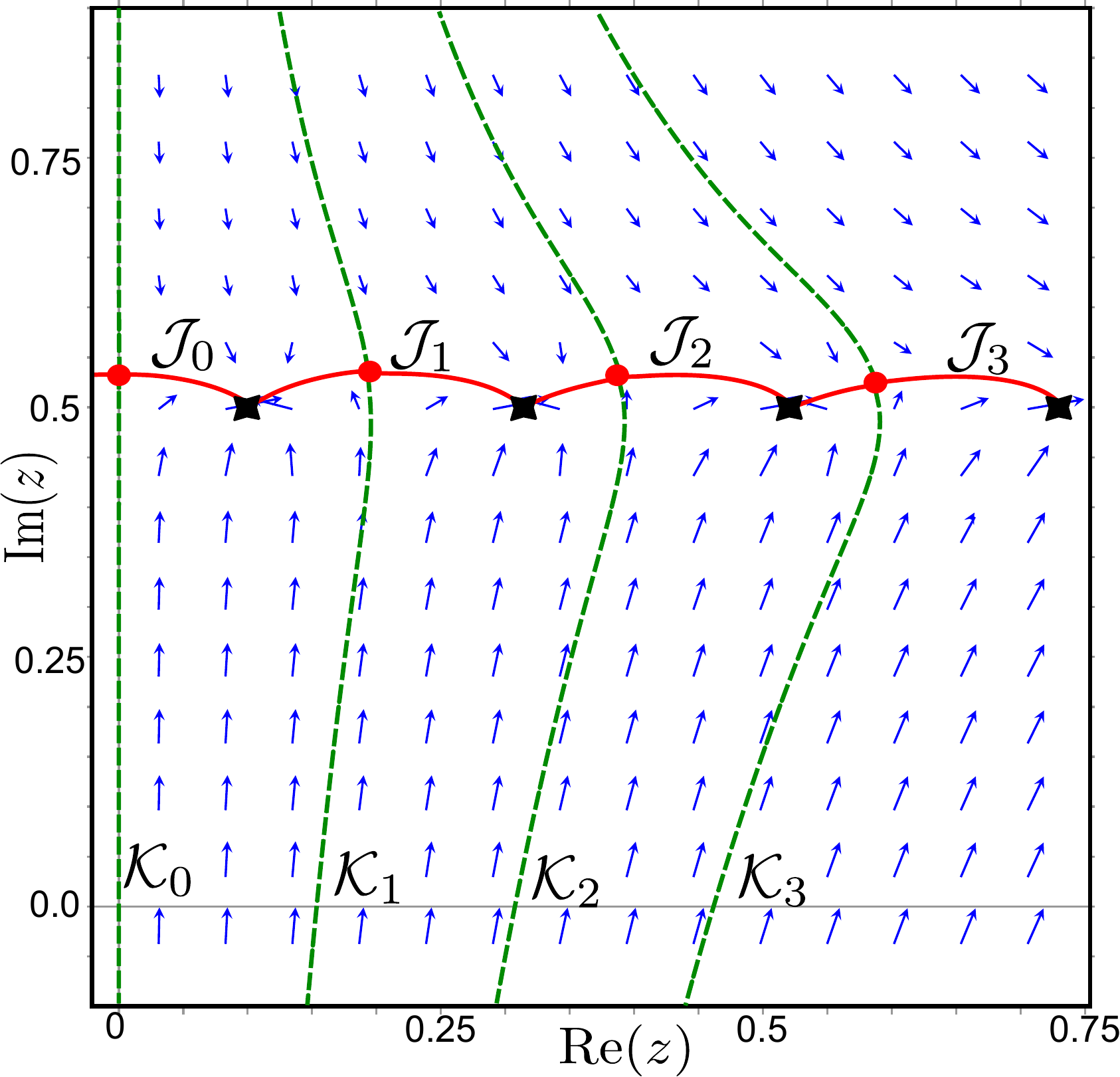}
}
\end{minipage}
\caption{
Behaviors of Morse's downward flow equation for $\beta U=30$, $U=1$, and $\mu/U=2$ (a) [$\mu/U=0$ (b)].
Star-shape black points show singular points of logarithm, and red blobs show complex saddle points $z_{\sigma}$. 
}
\label{fig:Flow_HubbardModel}
\end{figure}

Behaviors of the downward flow equation \eqref{Eq:Hubbard_DownwardFlow} are shown in Fig.~\ref{fig:Flow_HubbardModel} with $U=1$ and $\beta U=30$. 
Let us discuss the case when the chemical potential is sufficiently large ($\mu/U=2$). 
In this case, Morse's flow is shown in Fig.~\ref{fig:Flow_HubbardModel}~(a). Only one Lefschetz thimble $\mathcal{J}_*$ associated with $z_*$ contributes in the Lefschetz-thimble decomposition. 
Other dual thimbles shown with dashed green lines do not intersect with the original integration cycle $\mathbb{R}$ as in Fig.~\ref{fig:Flow_HubbardModel}~(a). 
Thus intersection numbers in Eq.~\eqref{Eq:Lefschetz_Thimble_Decomposition} vanish and we have 
\be
Z=\int_{\mathcal{J}_*}\diff z~\mathrm{e}^{-S(z)}. 
\ee 
Within the saddle-point approximation using the complex saddle point $z_*$, the number density is given as 
\be
n\simeq {-\im z_*\over U}=2, 
\ee 
and thus the saturation of fermions is well described. 
The exactly similar thing happens also for $\mu/U\lesssim -1/2$,  %$\mu/U\lesssim -1$, 
and then the saddle-point approximation gives $n\simeq -\im z_*/U=0$. 
This analysis explicitly shows why the sign problem is significantly weakened with an appropriate shift of integration variables for $\mu/U\lesssim -1/2$ and $\mu/U\gtrsim 3/2$,  as we have discussed in Sec.~\ref{sec:HubbardModel_pathintegral}. 

In the following, we concentrate on the case where the sign problem is severe, $-1/2\lesssim \mu/U\lesssim 3/2$, at low temperatures $\beta U\gtrsim 10$. 
In this parameter region, the typical behavior of the flow is shown in Fig.~\ref{fig:Flow_HubbardModel}~(b). 
All dual thimbles $\mathcal{K}_m$ intersect with the original integration cycle $\mathbb{R}$. 
This shows that all the saddle points $z_{m}$ contribute to the partition function, so that the interference among them may not be negligible. 
This interference requires a careful treatment of the semiclassical analysis in order to solve the sign problem. 
The big difference between Figs.~\ref{fig:Flow_HubbardModel}~(a) and (b) can be explained by the Stokes phenomena~\cite{Witten:2010cx}, which occur around $\mu/U\simeq 3/2$ and also around $\mu/U\simeq-1/2$. 
However, this is irrelevant to the non-analytic behavior of the number density, and then we stick to the analysis for $-1/2\lesssim \mu/U\lesssim 3/2$. 

Let us denote the classical action at the saddle point $z_{\sigma}$ as $S_{\sigma}(:=S(z_{\sigma}))$. 
Substituting the approximate expression \eqref{Eq:ApproximateSaddlePoints} into Eq.~\eqref{Eq:EffectiveAction_Hubbard}, we have  
\begin{eqnarray}
&S_0\simeq-{\beta U\over 2}\left(\frac{\mu}{U}+{1\over 2}\right)^2,
\label{Eq:ApproximateClassicalAction_OneThimble}\\
&\mathrm{Re}\,(S_m-S_0)\simeq {2\pi^2\over \beta U}m^2, 
\label{Eq:ApproximateClassicalActionRe}\\
&\mathrm{Im}\,S_m\simeq 2\pi m\left({\mu\over U}+\frac{1}{2}\right).
\label{Eq:ApproximateClassicalActionIm} 
\end{eqnarray}
For Eqs.~\eqref{Eq:ApproximateClassicalActionRe} and \eqref{Eq:ApproximateClassicalActionIm}, we have written down the $m$-dependent leading terms in terms of large $\beta U$. 
Equation \eqref{Eq:ApproximateClassicalActionRe} shows that subdominant thimbles $\mathcal{J}_m$ can be comparable with the dominant one $\mathcal{J}_0$ for $\beta U\gg 1$ so long as $|m|(\not=0)$ is not so large. 
According to Eq.~\eqref{Eq:ApproximateClassicalActionIm}, these different thimbles have different complex phases, and thus a contribution from one Lefschetz thimble is generally  canceled by other ones. 
Exactly at the half-filling, $\mu/U=1/2$, the complex phase is always an integer multiple of $2\pi \im$, and thus such a cancellation is absent. This gives an interpretation on the reason why the sign problem disappears at the half-filling from the viewpoint of Lefschetz-thimble approach. 

Let us compute the partition function $Z_{\mathrm{cl}}$ only using these semiclassical information: 
%Neglecting the higher order correction in the Gaussian fluctuation, let us simply compute
\be
Z_{\mathrm{cl}}:=\sum_{m=-\infty}^{\infty}\mathrm{e}^{-S_m}. 
\label{Eq:ApproximateSemiclassicalSummation}
\ee
In this formula, Gaussian fluctuation around $z_m$ is neglected because it only gives an unimportant overall factor at low temperatures. Indeed, within our approximation, $\p_z^2 S(z_m)$ is independent of $m$, and the result of the Gaussian integration slip by the summation. 
This summation \eqref{Eq:ApproximateSemiclassicalSummation} can be computed analytically by using the elliptic theta function as 
\begin{eqnarray}
Z_{\mathrm{cl}}&\simeq\mathrm{e}^{-S_0}\left(1+2\sum_{m=1}^{\infty}\cos 2\pi m\left({\mu\over U}+{1\over 2}\right) \mathrm{e}^{-{2\pi^2m^2/\beta U}}\right)\nonumber\\
&=\mathrm{e}^{-S_0}\theta_3\left(\pi\left({\mu\over U}+{1\over 2}\right),\mathrm{e}^{-2\pi^2/\beta U}\right).
\label{Eq:ApproximateSemiclassicalPartitionFunction}
\end{eqnarray}
Here we used an approximate expression of the saddle points, Eqs.~\eqref{Eq:ApproximateClassicalActionRe} and  \eqref{Eq:ApproximateClassicalActionIm}, to obtain Eq.~\eqref{Eq:ApproximateSemiclassicalPartitionFunction}. 
Using this result, in the limit $\beta\to \infty$, the number density is given as 
\be
n_{\mathrm{cl}}:={1\over \beta}{\p\over \p \mu}\ln Z_{\mathrm{cl}}
\to 
\left\{\begin{array}{cl}
2& (1<\mu/U<3/2),\\
1& (0<\mu/U<1),\\
0& (-1/2<\mu/U<0). 
\end{array}\right.
\label{eq:Number_Density_Semiclassical_T0}
\ee
This good agreement might be accidental to some extent, since we have neglected higher order contributions in the semiclassical analysis. 
This result still indicates the usefulness of the semiclassical approximation even when the sign problem is severe. 

\subsection{Lee--Yang zeros and the fermion spectrum}\label{sec:LeeYangHubbard}
Let us analyze the non-analytic behavior \eqref{eq:Number_Density_Semiclassical_T0} from the viewpoint of Lee--Yang zeros~\cite{PhysRev.87.404, PhysRev.87.410, Fodor:2001pe, Fodor:2004nz, Ejiri:2005ts, Stephanov:2006dn, Kratochvila:2006jx, Nagata:2012tc, Nagata:2014fra} in the semiclassical expression \eqref{Eq:ApproximateSemiclassicalPartitionFunction}. 
Using the infinite-product expression of the elliptic theta function,
\be
\theta_3(z,q)=\prod_{\ell=1}^{\infty}\left\{\left(1-q^{2\ell}\right)\left(1+2q^{2\ell-1}\cos(2z)+q^{4\ell-2}\right)\right\} . 
\ee
We find that the zeros of the partition function in the complex $\mu$ plane satisfy 
\be
\cos \left(2\pi{\mu\over U}\right)=\cosh \left({2\pi^2\over \beta U}(2\ell-1)\right) 
\ee
for some positive integer $\ell$. Since Eq.~\eqref{Eq:ApproximateSemiclassicalPartitionFunction} is originally derived as an approximate expression for $-1/2<\mu/U<3/2$, we restrict the real part of the complex chemical potential to $-U/2<\mathrm{Re}\,\mu<3U/2$. 
Then, the zeros of the semiclassical partition function are 
\be
\mu=\pm \im \pi T(2\ell-1),\; \mu=U\pm \im\pi T(2\ell-1).  
\ee
In the limit $T\to 0$, these zeros form straight lines crossing to the real axis at $\mu=0, U$. 
This behavior of zeros is a signal of the first-order phase transition in the context of the Lee--Yang zero analysis of a therodynamic system, and indeed we find the non-analytic jumps~\eqref{eq:Number_Density_Semiclassical_T0} of $n$ at $\mu=0, U$ in the limit of $T=0$ in the one-site Hubbard model. 

Let us apply the same analysis to the phase quenched partition function. We define the phase quenched semiclassical partition function by 
\be
Z'_{\mathrm{cl}}:=\sum_{m=-\infty}^{\infty}\Bigl|\mathrm{e}^{-S_m}\Bigr|=\mathrm{e}^{-S_0}\theta_3\left(0,\mathrm{e}^{-2\pi^2/\beta U}\right).
\ee
Note that this expression does not correspond to the semiclassical limit of Eq.~\eqref{Eq:QuenchedZ_woShift} because the saddle points at $-1/2<\mu/U<3/2$ do not lie on the original integration cycle $\mathbb{R}$. 
In the phase quenched approximation, the interference of complex phases $\mathrm{e}^{-\im\, \mathrm{Im}S_m}$ disappears, and then the $\mu$ dependence is given only by $\mathrm{e}^{-S_0}$. 
Lee--Yang zeros do not exist if $-U/2\ll \mathrm{Re}\,\mu\ll 3U/2$. 
Therefore, the only possibility to explain the non-analytic behavior in this phase quenched approximation originates from the Stokes jumps around $\mu/U\simeq -1/2,3/2$, which naturally explains the non-analytic but continuous change of $n$ (see Ref.~\cite{Kanazawa:2014qma}). 

In order to investigate this failure of the phase quenched approximation more deeply, 
we analyze the fermion spectrum \eqref{eq:FermionSpectrumHubbard} at complex saddle points. 
For simplicity, let us take the continuum limit at first with an appropriate normalization, 
\be
\lambda^{\mathrm{con}}_{\ell}(\varphi,\mu):=\lim_{N\to \infty}TN\lambda_\ell (\varphi,\mu)
=-\left\{(2\ell -1)\pi \im T+\im \varphi+\mu+{U\over 2}\right\}. 
\label{eq:OneSiteModel_fermion_spectrum}
\ee
The spectral representation of the number density at a background field $\varphi$ is given by 
\be
n(\varphi,\mu)=-2T\sum_{\ell=-\infty}^{\infty}{\mathrm{e^{\im0^+\ell}}\over \lambda^{\mathrm{con}}_{\ell}(\varphi,\mu)}={2\over \mathrm{e}^{-\beta(\im \varphi+\mu+U/2)}+1}. 
\ee
If the sign problem is mild enough, $\mu<-U/2$, then the eigenvalue at the saddle point $z_*=0$ is 
\be
\lambda^{\mathrm{con}}_{\ell}(z_*,\mu)\simeq-\left((2\ell -1)\pi \im T+\mu+{U\over 2}\right). 
\ee
Since $\mathrm{Re}\,\lambda^{\mathrm{con}}_{\ell}(z_*,\mu)$ cannot be zero, the chemical potential dependence of the number density is exponentially suppressed as, in the limit $T\to 0$, 
\be
n(z_*,\mu)\simeq {2\over \mathrm{e}^{-\beta(\mu+U/2)}+1}\simeq 0. 
\ee
However, for $\mu>-U/2$, $\mathrm{Re}\,\lambda_{\ell}^{\mathrm{con}}$ can be zero in the vicinity of saddle points, and indeed one finds 
\be
\lambda^{\mathrm{con}}_{\ell}(z_m,\mu)\simeq -\left((2(\ell+m) -1)\pi \im T-T\ln {{3\over2}U-\mu\over {U\over 2}+\mu}\right). 
\ee
Within this approximation, the number density becomes 
\be
n(z_m,\mu)= {2\over \mathrm{e}^{-\beta(\im z_m+\mu+U/2)}+1}\simeq {\mu\over U}+{1\over 2}. 
\label{Eq:NumberDensity_PhaseQuenched}
\ee
The near-zero fermionic mode provokes the fake onset of the number density at $\mu=-U/2$ again. 
From the viewpoint of the fermion spectrum at complex saddle points, nothing special happens in the vicinity of the correct transition points $\mu=0, U$. 
In the language of Lefschetz thimbles, this implies that their topological structure does not change at all around $\mu=0,U$, as we can see in Fig.~\ref{fig:Flow_HubbardModel}~(b). 

\subsection{Numerical results}
\label{sec:NumericalResults_OneSiteHubbard}

\begin{figure}[t]
\centering
\includegraphics[scale=1.2]{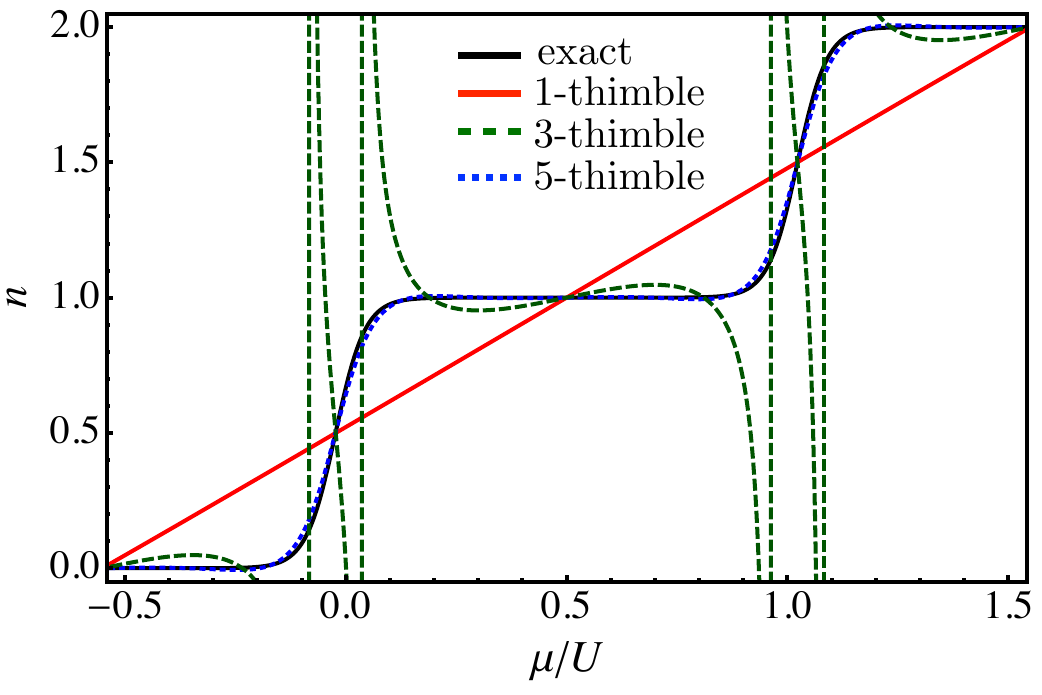}
\caption{
Behaviors of the number density $n=-\im\langle\varphi\rangle/U$ as a function of $\mu/U$ at $\beta U=30$. 
The solid black line shows the exact solution. 
Other lines, solid red one, dashed green one, and dotted blue one,
show the result of one-, three-, and five-thimble approximations, which integrate over $\mathcal{J}_0$, $\mathcal{J}_{0}\cup\mathcal{J}_{\pm 1}$, and $\mathcal{J}_{0}\cup\mathcal{J}_{\pm1}\cup\mathcal{J}_{\pm2}$, respectively. 
\label{fig:NumberDensityHubbardBeta30}
}
\end{figure}

Let us consider one-, three-, and five-thimble approximations in order to analyze the importance of the interference among multiple Lefschetz thimbles. 
We consider the case where the sign problem is severe, i.e., $-1/2\lesssim \mu/U\lesssim 3/2$ and $\beta U\gtrsim 10$. 
The $(2m+1)$-thimble approximation takes into account only $\mathcal{J}_0,\mathcal{J}_{\pm 1},\ldots,\mathcal{J}_{\pm m}$ in Fig.~\ref{fig:Flow_HubbardModel}~(b). The partition function~\eqref{Eq:ZeroModeIntegral_Hubbard} in this approximation reads 
\be
Z\simeq Z^{(2m+1)}:= \int_{\mathcal{J}_0}\diff z\;\mathrm{e}^{-S(z)}+\sum_{k=1}^{m}2\mathrm{Re}\int_{\mathcal{J}_{k}}\diff z\; \mathrm{e}^{-S(z)}. 
\label{Eq:2n+1ThimbleApproximation}
\ee
Each contribution of the summation in the second term comes from a pair $\mathcal{J}_k$ and $\mathcal{J}_{-k}$, and becomes manifestly real because of the reflection symmetry $z\mapsto -\overline{z}$. 
Since the Lefschetz thimble $\mathcal{J}_k$ is homologically equivalent to the interval $\im(\mu+U/2)+[(2k-1)\pi T,(2k+1)\pi T]$, the above integration \eqref{Eq:2n+1ThimbleApproximation} is reduced to  
\be
Z^{(2m+1)}= \int_{-(2m+1)\pi T}^{+(2m+1)\pi T}\diff x\; \mathrm{e}^{-S\left(x+\im(\mu+U/2)\right)}, 
\ee
with $x=z-\im (\mu+U/2)$. 
The result on the number density at $\beta U =30$ is shown in Fig.~\ref{fig:NumberDensityHubbardBeta30}, in which one-, three-, and five-thimble approximations are compared with the exact computation. 

The one-thimble approximation, which is shown with the solid red line in Fig.~\ref{fig:NumberDensityHubbardBeta30}, is an approximation to integrate only over the Lefschetz thimble $\mathcal{J}_0$. 
It almost gives the mean-field result, that is,
\be
n_{\mathrm{MF}}={\mu\over U}+{1\over 2}, 
\label{Eq:MeanFieldNumberDensity}
\ee
for $-1/2<\mu/U<3/2$. 
Figure~\ref{fig:NumberDensityHubbardBeta30} shows that the one-thimble approximation is not sufficient to describe the plateaus of the number density in the region $-1/2<\mu/U<3/2$. 
One can easily check that Eq.~\eqref{Eq:MeanFieldNumberDensity} is also obtained from the phase quenched approximation \eqref{Eq:NumberDensity_PhaseQuenched} after shifting the integration variable in Eq.~\eqref{Eq:ZeroModeIntegral_Hubbard} as $\varphi_{\mathrm{bg}}\mapsto \varphi_{\mathrm{bg}}+\im(\mu+U/2)$. 

The result of the three-thimble approximation provides us a useful lesson for an application of the Lefschetz-thimble approach to the sign problem. 
The number density diverges at several chemical potentials around a rapid crossover, although the result is improved around each plateau. 
Let us analyze this divergence by using the semiclassical analysis. 
For that purpose, we introduce the semiclassical partition function with the $(2m+1)$-thimble approximation by 
\be
Z_{\mathrm{cl}}^{(2m+1)}:=\sum_{k=-m}^{m}\mathrm{e}^{-S_k}. 
%\exp-S_k. 
\ee
In the three-thimble approximation, the semiclassical partition function approximately behaves as 
\be
Z_{\mathrm{cl}}^{(3)}\simeq \mathrm{e}^{-S_0}\left[1+2\mathrm{e}^{-{2\pi^2/ \beta U}} \cos 2\pi \left({\mu\over U}+\frac{1}{2}\right)\right].
\ee
Since $\beta U\gg 1$ and thus $\mathrm{e}^{-{2\pi^2/ \beta U}}\sim 1$, the second term is not suppressed, compared with the first term. This approximate partition function vanishes at some chemical potentials around $\mu/U=0$ and $1$ ($Z_{\mathrm{cl}}^{(3)}$ vanishes at $\mu/U\simeq\pm 0.04$, which is roughly consistent with Fig.~\ref{fig:NumberDensityHubbardBeta30}). 
This example clearly demonstrates that picking up a part of the Lefschetz-thimble decomposition can violate the physical requirement, such as the positivity condition on the thermodynamic quantities. 
In this instance, the incompressibility $(\partial n/\partial \mu)/n^2$ should not be negative.

According to Fig.~\ref{fig:NumberDensityHubbardBeta30}, we can conclude that five Lefschetz thimbles $\mathcal{J}_{0}\cup\mathcal{J}_{\pm1}\cup\mathcal{J}_{\pm2}$ are necessary in order to explain rapid jumps of the number density in terms of the chemical potential at a low temperature $\beta U=30$. 
In fact, the five-thimble approximation (dotted blue line) well reproduces the exact computation (solid black line) as seen in Fig.~\ref{fig:NumberDensityHubbardBeta30}. 
How many Lefschetz thimbles are necessary in general at a given lower temperature $\beta$? 
We propose the criterion to neglect the Lefschetz thimbles $\mathcal{J}_m$ for large $|m|$: 
\be
\left|{Z_{\mathrm{cl}}^{(2m+1)}-Z_{\mathrm{cl}}^{(2m-1)}\over Z_{\mathrm{cl}}^{(2m+1)}}\right|\ll 1. 
\ee
Assuming that the number of Lefschetz thimbles, $(2m+1)$, is sufficiently large, one can replace $Z_{\mathrm{cl}}^{(2m+1)}$ in the denominator by the semiclassical partition function $Z_{\mathrm{cl}}$. 
Since $\bigl|Z_{\mathrm{cl}}^{(2m+1)}-Z_{\mathrm{cl}}^{(2m-1)}\bigr|\le 2|\exp-S_m|$, the contribution from $\mathcal{J}_m$ can be negligible if 
\be
{2|\exp-S_m|\over Z_{\mathrm{cl}}} \lesssim \ve. 
\ee
Here $\ve$ is a controlling parameter of an error, and $\ve\ll 1$. 
Using the approximate results \eqref{Eq:ApproximateClassicalActionRe} and \eqref{Eq:ApproximateSemiclassicalPartitionFunction}, we can solve this inequality with respect to $m$: 
\bea
|m|\gtrsim \sqrt{-{\beta U\over 2\pi^2}\ln {\ve\over 2}\theta_3\left(\pi\left({\mu\over U}+{1\over 2}\right), \mathrm{e}^{-2\pi^2/\beta U}\right)}. 
\label{Eq:NecessaryNumbersOfThimbles}
\eea
This criterion gives different results depending on $\mu$, and thus let us first derive the strongest restriction. If the sign problem is severe, i.e., $\mu=0$, the elliptic theta function is exponentially small with respect to $\beta U$:
\be
\theta_3\left({\pi\over 2},\mathrm{e}^{-2\pi^2/\beta U}\right)\simeq \sqrt{2\beta U\over \pi}\mathrm{e}^{-\beta U/8}. 
\ee
Therefore, for reasonable $\ve$ such as $\ve\simeq 0.1$, the effect of $\ve$ is negligible and the criterion~\eqref{Eq:NecessaryNumbersOfThimbles} gives 
\be
|m|\gtrsim {\beta U\over 4\pi}
\label{Eq:NecessaryNumbersOfThimbles_Mu0}
\ee
in the limit of $\beta U\gg 1$. 
It means that we need at least $(2\lceil\beta U/4\pi\rceil +1)$ thimbles in order to describe the rapid crossover of the number density in the case of the one-site Hubbard model. 

Let us discuss this behavior from another point of view. Since $\beta$ becomes large, the relevant integration region becomes smaller as $|\mathrm{Re}\; z|\lesssim \sqrt{U/\beta}$. 
However, Fig.~\ref{fig:Flow_HubbardModel} shows that the length of each Lefschetz thimble $\mathcal{J}_k$ is proportional to $1/\beta$, and thus we need more Lefschetz thimbles in order to cover the relevant integration region. 
That number is clearly proportional to $\sqrt{\beta U}$, and is much smaller than Eq.~\eqref{Eq:NecessaryNumbersOfThimbles_Mu0}. 
Because of the interference of complex phases among Lefschetz thimbles, the semiclassical partition function becomes much smaller than that of the naive expectation if $\mu/U\sim 0,\, 1$. 
This indicates that it is much easier to reproduce the plateau in the vicinity of the half-filling than that of the rapid crossover. 
The necessary number of Lefschetz thimbles for each phenomenon is also largely different. 
In fact, for $\mu/U=1/2$, one can show that 
\be
{Z_{\mathrm{cl}}\over \mathrm{e}^{-S_0}}\simeq \theta_3\left(\pi, \mathrm{e}^{-2\pi^2/\beta U}\right)\simeq \sqrt{\beta U\over 2\pi}, 
\ee
and, roughly speaking, the criterion~\eqref{Eq:NecessaryNumbersOfThimbles}  gives 
\be
|m|\gtrsim \sqrt{{\beta U\over 2\pi^2}\ln{\sqrt{8\pi}\over \ve\sqrt{\beta U}}}
\label{Eq:NecessaryNumbersOfThimbles_Mu1/2}
\ee
for $\mathcal{J}_m$ being negligible with keeping a good accuracy only around the half filling. This is consistent with the result of the above heuristic argument. 
Now, we can understand that the large gap of these two estimates \eqref{Eq:NecessaryNumbersOfThimbles_Mu0} and \eqref{Eq:NecessaryNumbersOfThimbles_Mu1/2} comes from the sign problem in summing up Lefschetz thimbles, and the semiclassical analysis provides a reasonable rough indication. 

%---------------------------------------------------------------------------------
%Speculation
%---------------------------------------------------------------------------------
\section{Speculation on the Silver Blaze problem in finite-density QCD}\label{sec:speculation_silverblaze_qcd}

We studied the one-site Hubbard model as a toy model of the sign problem, and the Silver Blaze problem. We elucidate that the interference of complex phases among multiple saddle points is important to analyze these problems. 
In this section, we discuss the analogy of the Silver Blaze problem in the one-site Hubbard model with that in finite-density QCD. 

\begin{figure}[t]
\centering
\includegraphics[scale=0.58]{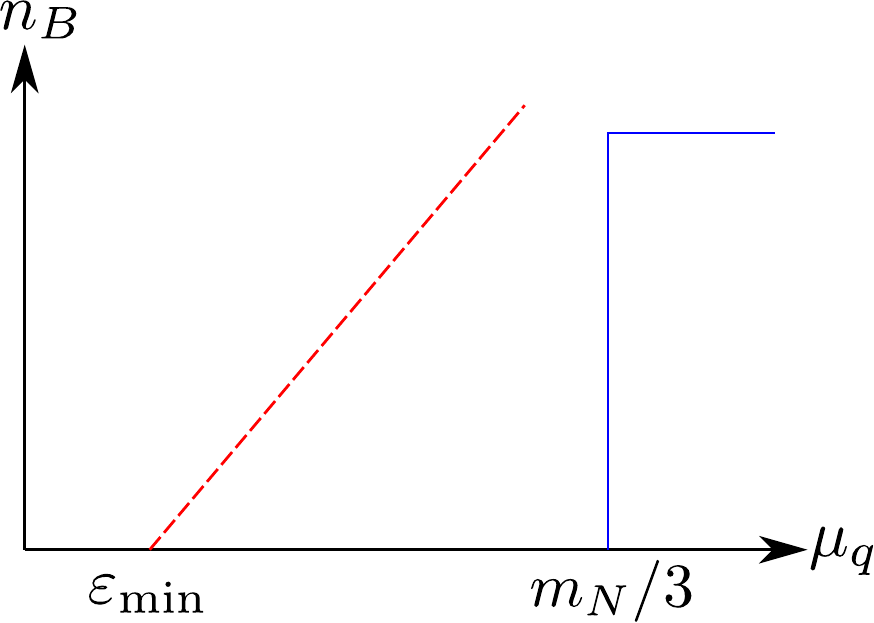}
\caption{
Schematic illustration of the Silver Blaze problem in the finite-density QCD. The baryon number density jumps at $\mu\simeq m_N/3$ (solid blue line).  Phase quenched theory shows the early onset of the baryon number density (dashed red line). 
}
\label{fig:schematic_qcd_silver_blaze}
\end{figure}

Let us pay attention to the behavior of the baryon number density $n_B$ for finite-density QCD at $T=0$. 
%In general, eigenvalues $\lambda_j(A,\mu_\mathrm{qk})$ of the fermion operator, $\gamma_4(\slashed{D}_A+m)-\mu_\mathrm{qk}$, depend on the quark chemical potential $\mu_\mathrm{qk}$, where $\slashed{D}_A=\gamma_{\nu}(\p_{\nu}-\im A_{\nu})$.  
In general, the quark determinant, $\mathrm{Det}\left[\gamma_4(\slashed{D}_A+m)-\mu_\mathrm{qk}\right]$, will have the $\mu_\mathrm{qk}$-dependence, which naively means that the baryon number density arises for any chemical potentials. 
%Therefore, the quark determinant, $\mathrm{Det}\left[\gamma_4(\slashed{D}_A+m)-\mu_\mathrm{qk}\right]=\prod_{\color{red}\ell} \lambda_{\color{red}\ell}(A,\mu_\mathrm{qk})$, will have the $\mu_\mathrm{qk}$-dependence, which naively means that the baryon number density arises for any chemical potentials. 
However, since we know that the lightest baryon in the QCD spectrum is nucleon, which is a composite particle of three quarks, the baryon number density $n_B$ must be zero for any $3\mu_\mathrm{qk}<m_N-B\simeq 923~\mathrm{MeV}$ ($m_N$ is the nucleon mass and $B$ is the nuclear binding energy).  
The $\mu_\mathrm{qk}$-dependence of the QCD partition function must vanish for $\mu_\mathrm{qk}\lesssim m_N/3$ at $T=0$, which empirical fact is highly nontrivial from the viewpoint of the path integral of lattice QCD as discussed above (Fig.~\ref{fig:schematic_qcd_silver_blaze} illustrates this situation schematically). 
Understanding this behavior of $n_B$ must be a key first step to correctly perform the lattice QCD simulation at arbitrary temperatures and baryon chemical potentials. In Ref.~\cite{Cohen:2003kd}, this is named the baryon Silver Blaze problem after the famous story of Sherlock Holmes. 

So far, the baryon Silver Blaze problem has been well understood only for $\mu_\mathrm{qk}< m_{\pi}/2$ ($m_{\pi}\simeq140~\mathrm{MeV}$ is the pion mass)~\cite{Cohen:2003kd, Adams:2004yy, Nagata:2012tc}. 
In general, the eigenvalues of $\gamma_4(\slashed{D}_A+m)-\mu_\mathrm{qk}$ can be expressed as $\lambda_{(j,n)}(A,\mu_\mathrm{qk})=\varepsilon_j(A)-\mu_\mathrm{qk}-\mathrm{i}\phi_j(A) +\mathrm{i}(2n+1)\pi T$ with $-\pi T<\phi_j\le\pi T$~\cite{Cohen:2003kd}. 
Therefore, $\mathrm{Det}\left[\gamma_4(\slashed{D}_A+m)-\mu_\mathrm{qk}\right]=\prod_j\prod_n \lambda_{(j,n)}$. 
The product over $n$ can be taken in the exactly same way as we usually do for that over Matsubara frequencies, and we find that it is proportional to $\left(\mathrm{e}^{\beta(\ve_j-\im\phi_j-\mu_{\mathrm{qk}})/2} +\mathrm{e}^{-\beta(\ve_j-\im\phi_j-\mu_{\mathrm{qk}})/2}\right)$ for each label $j$. 
The fermion determinant becomes
\bea
&\mathrm{Det}\left[\gamma_4(\slashed{D}_{A}+m)-\mu_\mathrm{qk}  \right] \nonumber\\
&=\mathcal{N} \prod_j\left(\mathrm{e}^{\beta(\ve_j-\im\phi_j-\mu_{\mathrm{qk}})/2} +\mathrm{e}^{-\beta(\ve_j-\im\phi_j-\mu_{\mathrm{qk}})/2}\right)\nonumber\\
&=\mathcal{N} \,\mathrm{e}^{-\beta \mu_\mathrm{qk}\, \mathrm{Tr}(1)/2} \prod_j\left(1+\e^{-\beta(\varepsilon_j-\mu_\mathrm{qk})}\e^{\mathrm{i}\beta\phi_j}\right)\nonumber\\
&=\mathcal{N} \prod_{\ve_j>0}\mathrm{e}^{\beta (\ve_j+\im\phi_j)}\left(1+\mathrm{e}^{-\beta(\ve_j-\mu_\mathrm{qk}-\im \phi_j)}\right)\left(1+\mathrm{e}^{-\beta(\ve_j+\mu_\mathrm{qk}+\im \phi_j)}\right), 
\label{eq:DetQCD}
\eea
where $\mathcal{N}$ is the normalization factor (see Ref.~\cite{Adams:2004yy} for its more rigorous derivation using the zeta-function regularization). 
Its $\mu_\mathrm{qk}$-dependence at sufficiently low temperatures and at $\mu_\mathrm{qk}>0$ becomes crystal clear according to the following expression \cite{Cohen:2003kd, Adams:2004yy}: 
\be
{\mathrm{Det}\left[\gamma_4(\slashed{D}_{A}+m)-\mu_\mathrm{qk}  \right]\over \mathrm{Det}\left[\gamma_4(\slashed{D}_{A}+m)\right]}\simeq  \prod_{0<\ve_j(A)<\mu_\mathrm{qk}}\left(1+\mathrm{e}^{-\beta(\ve_j(A)-\mu_\mathrm{qk})}\e^{\mathrm{i}\beta\phi_j(A)}\right). 
\ee
The quark determinant becomes independent of $\mu_\mathrm{qk}$ at $T=0$ if $\mu_\mathrm{qk}$ is smaller than the minimum of $\ve_j(A)(>0)$. For statistically significant gauge fields, it is confirmed that $\mathrm{min}(\ve_j(A))=m_{\pi}/2$ \cite{Cohen:2003kd, Adams:2004yy, Nagata:2012tc}, and thus $n_B=0$ for $\mu_{\mathrm{qk}}<m_{\pi}/2$. 
This also means that the Lefschetz-thimble decomposition of lattice QCD at $\mu_\mathrm{qk}<m_{\pi}/2$ and at sufficiently low temperatures is identical to the original integration cycle up to an exponentially small correction because the quark determinant at $\mu_{\mathrm{qk}}=0$ is positive.

In contrast, as computed in Eq.~(\ref{Eq:FermionicDeterminantHubbard}), the fermion determinant in the one-site Hubbard model is 
\be
\mathrm{Det}\left[\p_{\tau}-\left(\im \varphi(\tau)+\mu+{U\over 2}\right)\right]
=\left(1+\mathrm{e}^{-\beta(-U/2-\mu)}\mathrm{e}^{\im \beta\varphi_{\mathrm{bg}}}\right)^2. 
\label{eq:DetHubbard}
\ee
We can ask the same question: How can we show $n=0$ for $\mu<0$ in the limit $\beta=\infty$ by using the path-integral expression when the fermion determinant depends on $\mu$? 
This Silver Blaze problem for $\mu<-U/2$ can be solved in the same manner as the baryon Silver Blaze problem at $\mu_{\mathrm{qk}}<m_{\pi}/2$. At $\mu<-U/2$, the fermion determinant becomes independent of $\mu$ as $T\to 0$. 
This also explains why the Lefschetz thimble $\mathcal{J}_*$ at $\mu<-U/2$ is almost identical to the original integration cycle $\mathbb{R}$ (The similar discussion shows the Lefschetz thimble $\mathcal{J}_*$ at $\mu>3U/2$ is almost identical to the line with $\mathrm{Im}(z)=2$).

For $-U/2<\mu<0$, the $\mu$-dependence of the fermion determinant (\ref{eq:DetHubbard}) becomes exponentially large, and thus the fact that $n=0$ at $T=0$ is still veiled. 
At $\mu\simeq -U/2$, zeros of the fermion determinant (star-shape black points in Fig.~\ref{fig:Flow_HubbardModel}) come close to the real axis, which is almost identical to the Lefschetz thimble $\mathcal{J}_*$ for $\mu<-U/2$. 
In the case of the one-site Hubbard model, this triggers Stokes jumps and the original integration cycle is decomposed into multiple Lefschetz thimbles for $\mu>-U/2$. 
One can no longer replace the expectation value of the number density by the number density at a complex saddle point. 
The significance of interference among multiple Lefschetz thimbles is identified in order to explain not only the rapid jumps of $n$ but also the $\mu$-independence of $n$ ($n=0$) for $-U/2<\mu<0$ at the zero temperature. 
Furthermore, if the number of Lefschetz thimbles is insufficient, the thermodynamic stability is violated in the one-site Hubbard model. 

We speculate that this decomposition of Lefschetz thimbles and accompanying interference of complex phases also play an important role in finite-density QCD, in particular, in the Baryon Silver Blaze problem and in the sign problem beyond half of the pion mass. 
For $\mu_{\mathrm{qk}}>m_{\pi}/2$, the quark determinant becomes dependent on $\mu_{\mathrm{qk}}$ and highly oscillatory on the statistically significant domain of real gauge fields. 
This means that zeros of the quark determinant come close to that domain. It would be natural to conclude that the statistically dominant region of the original integration cycle starts to be decomposed into multiple Lefschetz thimbles around $\mu_{\mathrm{qk}}\simeq m_{\pi}/2$.  
The complex phase of the fermion determinant becomes ill-defined at its zeros. Thus, the constant-phase condition of the integrand on a Lefschetz thimble would be hard to be satisfied without such decomposition, when zeros of the fermion determinant are located in the vicinity of the Lefschetz thimble. 
Once the original path integral is decomposed into multiple thimbles, we might need to carefully take their interference into account to explain the $\mu_{\mathrm{qk}}$-independence of $n_B$ until $\mu_{\mathrm{qk}}$ exceeds $(m_N-B)/3$.

Because of the complexity of QCD, we cannot show these statements rigorously so far. Future study of QCD-like models based on Lefschetz thimbles and justifying our speculation will be crucial to develop our understanding on the baryon Silver Blaze problem and on the sign problem of the finite-density QCD beyond half of the pion mass. 

%---------------------------------------------------------------------------------
%Summary
%---------------------------------------------------------------------------------

\section{Summary}\label{sec:summary}

One-site repulsive Hubbard model shows a non-analytic behavior by changing the chemical potential at the zero temperature $\beta=\infty$. 
This model is obtained by taking the strong coupling limit, and thus our analysis of Lefschetz thimbles forms basics of studying the strongly-coupled repulsive Hubbard model using the path integral. 
Since the number eigenstate can solve this model exactly and its path integral expression has the sign problem, this is a good testing ground to analyze the structure of the sign problem in various solutions. 
Furthermore, the non-analytic behavior of the number density looks quite similar to the Silver Blaze problem in finite-density QCD, and thus the understanding of this model from the functional integral itself is an important task.

In this paper, the sign problem in the Lefschetz-thimble approach is carefully studied. 
Topological structures of the Lefschetz thimbles are analyzed based on the semiclassical analysis, and their relation to the fermion spectrum is investigated.
If one picks up only the most relevant Lefschetz thimble, the mean-field approximation is almost recovered. 
The significance of interference among multiple Lefschetz thimbles is identified in order to explain the non-analytic behavior. 
As we lower the temperature, there exist regions where the necessary number of Lefschetz thimbles increases linearly in $\beta U$. 
On the other hand, in order to explain the Silver Blaze like behavior, that number only grows as $\sqrt{\beta U}$. 

If the number of Lefschetz thimbles is insufficient, we found not only large discrepancies from the exact result, but also thermodynamic instabilities due to artifact of the approximation. 
This means that the Lefschetz-thimble decomposition does not manifestly ensure the thermodynamic stability, or the positivity condition on the partition function, although its reality is always true with a reasonable condition \cite{Tanizaki:2015pua}. 
Developing our understanding on the stability from the viewpoint of Lefschetz thimbles is important.

Since the one-site Hubbard model is quite a simple model, it is not clear whether interference of multiple Lefschetz thimbles is relevant for studying thermodynamic systems with complicated interactions. 
Nevertheless, we discuss the speculation on the baryon Silver Blaze problem of finite-density QCD based on its mathematical similarity to the one-site Hubbard model. We conjecture that the statistically dominant gauge-field configurations are decomposed into multiple Lefschetz thimbles when the quark chemical potential exceeds half of the pion mass. 
To justify our conjecture on finite-density QCD and QCD-like models is significantly important to solve the baryon Silver Blaze problem and to numerically simulate the finite-density QCD  beyond the half of the pion mass.

In order to understand strongly-correlated electron systems with the sign problem, applying the Lefschetz-thimble method to the two-site Hubbard model is a nice straightforward extension of this study. This enables us to take into account the effect of the hopping term. 
If the atomic potential is sufficiently strong, localization of fermions is a natural scenario and thus two-site models will become benchmark. 
We believe that semiclassical analysis with complex saddle points provides us a deeper understanding of the sign problem, and developing that technique in general cases is an important future study. 

% If you have acknowledgments, this puts in the proper section head.
\ack
% put your acknowledgments here.
Y.T. is supported by Grants-in-Aid for the fellowship of Japan Society for the Promotion of Science (JSPS) (No.25-6615). 
Y.H. is partially supported by JSPS KAKENHI Grants Numbers 15H03652.
This work was partially supported by the RIKEN interdisciplinary Theoretical Science (iTHES) project, and by the Program for Leading Graduate Schools of Ministry of Education, Culture, Sports, Science, and Technology (MEXT), Japan.

\appendix

\bibliography{./lefschetz,./ref_hubbard}
\end{document}